\RequirePackage{amsmath}
\documentclass[12pt]{iopart}
\usepackage{enumerate}
\usepackage{bbm}
\usepackage{amsmath}
\pdfoutput=1
\usepackage{braket}
\usepackage[paperwidth=210mm,paperheight=297mm,centering,hmargin=2cm,vmargin=2.5cm]{geometry}
\usepackage{relsize}
\usepackage[pdftex]{graphicx}
\usepackage[colorlinks=true,citecolor=blue]{hyperref}
\usepackage{cite}
\usepackage{mathrsfs}
\usepackage{array}
\usepackage{caption}

\usepackage{hyperref}
\usepackage{color}
\usepackage{graphicx,epsfig}
\usepackage{amssymb}
\usepackage[normalem]{ulem}

\newcommand{\ssp}{0.6ex}

\renewcommand{\d}{\text{d}}
\renewcommand{\dot}[1]{\overset{.}{#1}}

\newcommand{\trace}{{\rm Tr \,} }

\newcommand{\eps}{\epsilon}
\newcommand{\tpm}{\text{tpm}}
\newcommand{\MH}{\text{MH}}

\newcommand{\back}{\text{back}}
\def\be{\begin{eqnarray}}
\def\ee{\end{eqnarray}}

\newcommand{\uz}{^{(0)}}
\newcommand{\ut}{^{(\tau)}}
\newcommand{\uj}{^{(j)}}
\newcommand{\ujo}{^{(j-1)}}
\newcommand{\uk}{^{(K)}}
\newcommand{\uko}{^{(K-1)}}
\newcommand{\ujp}{^{(j+1)}}
\newcommand{\uo}{^{(1)}}
\newcommand{\ui}{^{(i)}}

\begin{document}

\title{Time-reversal symmetric work distributions for closed quantum dynamics in the histories framework \quad}

\author{Harry J. D. Miller and Janet Anders}

\address{Department of Physics and Astronomy, University of Exeter, Stocker Road, Exeter EX4 4QL, United Kingdom.}
\eads{hm419@exeter.ac.uk}


\begin{abstract}
A central topic in the emerging field of quantum thermodynamics is the definition of thermodynamic work in the quantum regime. One widely used solution is to define work for a closed system undergoing non-equilibrium dynamics according to the two-point energy measurement scheme. However, due to the invasive nature of measurement the two-point quantum work probability distribution cannot describe the statistics of energy change from the perspective of the system alone. We here introduce the quantum histories framework as a method to characterise the thermodynamic properties of the \emph{unmeasured}, closed dynamics. Constructing continuous power operator trajectories allows us to derive an alternative quantum work distribution for closed quantum dynamics that fulfils energy conservation and is time-reversal symmetric. This opens the possibility to compare the measured work with the unmeasured work, contrasting with the classical situation where measurement does not affect the work statistics. We find that the work distribution of the unmeasured dynamics leads to deviations from the classical Jarzynski equality and can have negative values highlighting distinctly non-classical features of quantum work.  

\end{abstract}
 
\maketitle

\section{Introduction}

Einstein's enquiry to Bohr if  ``the moon does not exist if nobody is looking at it''  questions the indeterminate nature of a quantum {\it state} when it is not measured \cite{Mermin1985}. Over the last century quantum mechanics' probabilistic predictions have proven unshakeably successful despite their counterintuitive philosophical implications. While quantum theory is inherently indeterministic, in classical statistical mechanics stochastic fluctuations arise due to ignorance about the state of the system. Classical statistical physics has elucidated that the macroscopic thermodynamic work, $\langle w \rangle$, can be considered as an average over fluctuating work values, $w$, assigned to individual trajectories in phase space taken by the system in each realisation of an experiment \cite{Jarzynski1997d,Crooks,Kawai2007a,Seifert2012}. Today research aims to link both quantum and classical stochasticity into a single quantum thermodynamic framework \cite{Vinjanampathy2015a}. Defining fluctuating work for quantum processes, and with it a work distribution $p(w)$ that characterises its statistical fluctuations, has been the subject of intense debate \cite{Allahverdyan2005a,Talkner2007c,Esposito2009,Allahverdyan2014c,Jarzynski2015a,Solinas2015,Alonso2015a,Deffner2016a,Perarnau-Llobet2016}. 

For closed quantum systems undergoing unitary dynamics a fluctuating work value has been identified within the Two-Point Measurement approach (TPM) \cite{Talkner2007c,Esposito2009,Hanggi2015}. Here the energy of the system is measured at the beginning and end of the system's evolution and the work, $w$, is taken to be the difference between these two measured energies. The two-point definition of work has been used to derive quantum fluctuation theorems \cite{Talkner2007c,Talkner2008b,Talkner2007a,Esposito2009,Kafri2012b,Hanggi2015}, which have been tested experimentally \cite{Batalhao2014,an2015}. The TPM scheme has also been generalised to open quantum systems \cite{Deffner2011a,Goold2015b,Talkner2009}. The choice of the TPM work definition has been supported by showing its direct correspondence with the classical definition of work in the semi-classical limit \cite{Jarzynski2015a}. 

However, as discussed in section \ref{sec:1law+time}, the TPM work distribution does not posses two properties that are generally associated with {\it closed} dynamics: (i) energy conservation requiring that the average work is exactly the internal energy change experienced by the system and (ii) the symmetry of the work distribution under time-reversal. The underlying cause for these two deviations is that, unlike in classical dynamics, the two energy measurements made in the TPM scheme affect the system dynamics in an irreversible way and make the system open \cite{Kammerlander2016}. 

To provide insight into how the measurement affects the work statistics we here derive an alternative quantum work distribution, $p(w)$, that characterises the {\it closed, unmeasured} quantum dynamics and fulfils the two physical properties (i) and (ii). The benefit of constructing this unmeasured $p(w)$ is that it allows the comparison with the TPM distribution for the measured dynamics. This comparison elucidates the influence of measurement on the statistics of work. Using the histories approach to quantum mechanics \cite{Griffiths1984,Gell-Mann1993a,HALLIWELL1995,Goldstein1995,Hartle2004} we construct the work distribution function associated with the continuous dynamics of the system in the absence of any experimental monitoring.  
Instead operator trajectories are introduced that describe the evolution in time through the set of instantaneous eigenvalues of a time-dependent power operator \cite{Solinas2013,PrasannaVenkatesh2015a}. We follow Goldstein and Page \cite{Goldstein1995} in assigning linear probabilities to general interfering trajectories, or \emph{histories}. This contrasts with other approaches, such as \cite{Subas2012,Li2013,PrasannaVenkatesh2015a}, that are restricted to evolutions that have no interference between histories. We find that the resulting work distribution is a quasi-probability distribution, implying that for general closed quantum dynamics fluctuating work cannot be interpreted as a measurable system property, while moments of the work distribution are well-defined and measurable. 

Thus while in classical physics work is associated with a system undergoing closed dynamics irrespective of whether the system is observed or not, in the quantum regime work is an indeterminate property of the non-equilibrium {\it process}. Measurements affect the quantum work distribution in a similar way as measurement collapses a quantum state. For commuting initial and final Hamiltonians of the dynamics we show that the moments of the quasi-probability work distribution reduce to the TPM values. 

The paper is organised as follows: in Section~\ref{sec:1law+time} we recap the TPM scheme and the work distribution requirements (i) and (ii). The histories formulation of quantum mechanics is  introduced in Section~\ref{sec:histories}. In Section~\ref{sec:pow} we use this framework to construct multi time-step trajectories for a quantum system in analogy to the classical phase space trajectories in statistical mechanics. We then derive the time-reversal symmetric quantum work distribution $p(w)$ and discuss its properties, exposing clear quantum signatures. In Section~\ref{sec:comparison} we compare $p(w)$ with a previously proposed power-based work distribution \cite{Li2013, PrasannaVenkatesh2015a}. A qubit example evidencing the negativity of $p(w)$ is discussed in Section~\ref{sec:example}. In Section~\ref{sec:MH} we show that the TPM work distribution can be recovered as a special case in the histories framework. Table~\ref{tab:compare} summarises the differing work statistics of the new time-reversal symmetric quasi-probability work distribution and known work distributions.  We summarise the results and open questions in Section~\ref{sec:discussion}.

\section{Energy conservation, time-reversal symmetry and the two-point measurement work distribution} \label{sec:1law+time}

We will begin by formally introducing conditions (i) and (ii) along with the TPM work distribution and show that in general the conditions fail for the TPM work.

{\bf Energy conservation.} In quantum mechanics the internal energy change, $\Delta U$, of a system starting with state $\rho$ and Hamiltonian $H(0)$ and ending with state $\rho(\tau)$ and Hamiltonian $H(\tau)$ in Schr\"odinger picture is universally identified as 
\be 
	\Delta U &:=&\trace[ H(\tau) \, \rho (\tau) - H (0) \, \rho ].
\ee
When the dynamics is \emph{closed}, i.e. the state evolves unitarily according to $\rho (\tau) = V(\tau) \, \rho \, V^{\dag} (\tau)$. The thermodynamic work done on a system is associated entirely with the energy change experienced by the system,
\be \label{1law}
	\langle w\rangle = \Delta U = \trace[(H_H(\tau) - H (0))  \, \rho ],
\ee
where we have changed into the Heisenberg picture, $H_H(\tau) = V^{\dag}({\tau}) \, H (\tau) \, V({\tau})$. This equation can be interpreted as an expression of the first law of thermodynamics for a closed system with no heat dissipation. 

Classical non-equilibrium statistical physics and stochastic thermodynamics have shown that the work $\langle w\rangle$ is to be understood as the \emph{average} over the fluctuating work associated with various trajectories the system undergoes in phase space \cite{Jarzynski1997d, Kawai2007a, Jarzynski2015a}. The measured fluctuating work values for each classical trajectory can be recast into likelihood for a specific work value, $w$, and hence a probability distribution for work, $p(w)$. This probability distribution allows the calculation of the mean work and higher work moments, 
\be \label{eq:avgs}
	\langle w^{m} \rangle = \int  w^{m} \, p(w) \, \d w .
\ee

\medskip

{\bf Time-reversal symmetry.}  A central concept in statistical physics is the symmetry of the dynamics under time-reversal. Physically one expects that a \emph{closed} system undergoing unitary dynamics has a work distribution that is symmetric under time-reversal, i.e. 
\be
	p_{\back}(w) = p(-w) .
\ee
Consequently the moments, Eq.~(\ref{eq:avgs}), would be anti-symmetric (odd powers) or symmetric (even). For example, in a closed system the average change in energy in a reverse process is expected to be the negative of the energy change experienced in the forward process. 

\medskip

{\bf TPM scheme.} 
The extension of thermodynamics to incooperate quantum properties requires the extension of the work concept to general quantum dynamics. However, the definition of fluctuating work, $w$, for the simplest case of a \emph{closed} quantum system has been subject to intensive discussion \cite{Allahverdyan2005a,Talkner2007c,Solinas2013,Allahverdyan2014c,Jarzynski2015a,Hanggi2015,Deffner2016a}. 
The current standard approach is to define the fluctuating work with the TPM scheme \cite{Talkner2007c,Esposito2009,Hanggi2015}, where a system's energy is measured twice: once before the start of the evolution at time $0$ with Hamiltonian $H(0)$, and then after the evolution with unitary $V(\tau)$ at time $\tau$ with Hamiltonian $H(\tau)$. In the TPM scheme the fluctuating work is  defined as the difference of the \emph{observed} energies, 
\be \label{eq:tpm}
	w_{n, m}=\eps_{m}\ut -\eps_{n}\uz, 
\ee
where $\eps_{n}\uz$ and $\eps_{m}\ut$ are the energy eigenvalues of the initial and final Hamiltonian, respectively, in either Schr\"odinger or Heisenberg picture. The joint probability associated with measuring these two energy eigenvalues is \cite{Esposito2009} 
\be \label{eq:jointtpm}
	p^{\tpm}_{n, m} = \trace[ \Pi_{m}\ut \,  \Pi_{n}\uz  \, \rho \,  \Pi_{n}\uz \, \Pi_{m}\ut ], 
\ee
where the system's state is $\rho$ and we have written the projectors onto the final energy eigenstates, $\Pi_{m}\ut$, in Heisenberg picture, i.e. $H_H (\tau)  =\sum_{m=1}^d \eps_{m}\ut \, \Pi_m\ut$. By summing over all energetic differences that give the same work value one obtains the quantum work probability distribution for the TPM scheme \cite{Talkner2007c,Esposito2009,Jarzynski2015a}
\be \label{eq:tpmw}
	p^{\tpm} (w) =  \sum_{n, m} \, \delta[w-(\eps_{m}\ut-\eps_{n}\uz)] \, p^{\tpm}_{n, m}.
\ee
However, as we will now see for some initial states $\rho$ and some evolutions $H(t)$ this work distribution does not to satisfy the energy conservation requirement and time-reversal symmetry expected to hold for closed systems. This is due to the invasiveness of quantum measurements.

Using Eq.~(\ref{eq:avgs}) and the quantum work distribution $p^{\tpm}(w)$ one finds that the average work in the TPM scheme is 
\be	 \label{eq:tpmwavg}
	\langle w \rangle_{p^{\tpm}(w)}=\trace[ H_H (\tau) \, \eta ] - \trace[H (0) \rho],
\ee 
where $\eta=\sum_{n}\Pi_{n}\uz \, \rho \, \Pi_{n}\uz$ is the state of the system projected into the initial energy basis. Importantly, the expectation value of the final Hamiltonian in state $\eta$ differs in general from the corresponding value for $\rho$. Thus for initial states that have coherences in the initial energy basis  \cite{Allahverdyan2014c,Solinas2015, Kammerlander2016,Elouard2015} the TPM work distribution $p^{\tpm}(w)$ cannot describe the total conserved average fluctuations in energy given by Eq.~(\ref{1law}).

In the TPM scheme the work distribution for the time-reversed process in Schr\"odinger picture is determined by taking the evolved state $\rho (\tau)$, measuring the final Hamiltonian $H(\tau)$, applying the reverse unitary $V^{\dag}$ and finally measuring the initial Hamiltonian $H(0)$. The associated time-reversed joint probability for two successive energy outcomes is given by $p^{\tpm}_{m,n \, \back} = \trace[ \Pi_{n}\uz \,  \Pi_{m}\ut  \, \rho \,  \Pi_{m}\ut \, \Pi_{n}\uz ]$. This expression is not symmetric to $p^{\tpm}_{n,m}$ given in Eq.~(\ref{eq:jointtpm}) unless $[\rho, H (0)]=0$ or $[H_H (\tau), H (0)]=0$. \cite{Johansen2007}. Therefore the TPM work distribution is not generally time-reversal symmetric, i.e.
\be
	p^{\tpm}_{\back}(- w) \neq p^{\tpm}(w),
\ee
reflecting the fact that the quantum measurements performed act as a source of irreversibility \cite{Yi2013b,Elouard2015,Lostaglio2015b}.

This contrasts with the classical notion of work which is independent of whether the system is measured or not. Here we wish to identify an alternative quantum work distribution $p(w)$ valid for universal unmeasured, closed quantum dynamics that remains consistent with the two natural conditions (i) and (ii). To achieve this we will make use of the histories framework for quantum mechanics, which we introduce in the following section.

\section{Histories for closed quantum systems}\label{sec:histories}

The quantum histories approach describes closed quantum systems and makes predictions equivalent to standard quantum mechanics \cite{Griffiths1984,HALLIWELL1995,Goldstein1995,Hartle2004,Hartle2008,Gell-Mann2012}. This approach allows one to assign a probability-like measure to quantum trajectories, also called histories, that extend the classical concept of trajectories in phase space. Here we introduce the histories approach which we will use in Section \ref{sec:pow} to derive a distribution function for quantum work under closed non-equilibrium dynamics. 

Let $\rho$ be the initial state of a system in a Hilbert space ${\cal H}$ of finite dimension $d$.  We consider the time-interval $[0, \tau]$ which is discretized into a set of $K$ time steps of duration $\Delta t = \tau / K$, with $t_{j}= j \, \Delta t$ and $j=\lbrace 0,1,..., K \rbrace$. The system is governed by a time-dependent Hamiltonian $H(t)$, which is decomposed into a discrete sequence in time $H\uj := H(t_j)$ and which generates a sequence of unitaries $V\uj = e^{- i {H\uj \Delta t}} \, V\ujo$ with $V\uz = \mathbbm{1}$ and $V\uk = V({\tau})$. Any Hermitian observable $X (t)$ discretized for time steps  $j$ can then be transformed into the Heisenberg picture as $X_H\uj = {V\uj}^{\dag} \, X\uj \, V\uj $. Throughout the remainder of the paper we will work in the Heisenberg picture unless stated otherwise. This means that the state remains $\rho$ at all times.  For each time step $j$ the operator $X_H\uj$ can be spectrally decomposed, $X_H\uj = \sum_{n=1}^d x_n\uj \, P_n\uj$, with orthonormal projectors $P_n\uj$ and corresponding  eigenvalues $ x_n\uj$ for $n=1, ..., d$. Throughout the paper we will assume non-degenerate eigenvalues and choose $\hbar =1$.

\begin{figure}[t]
\centering
\includegraphics[scale=0.40]{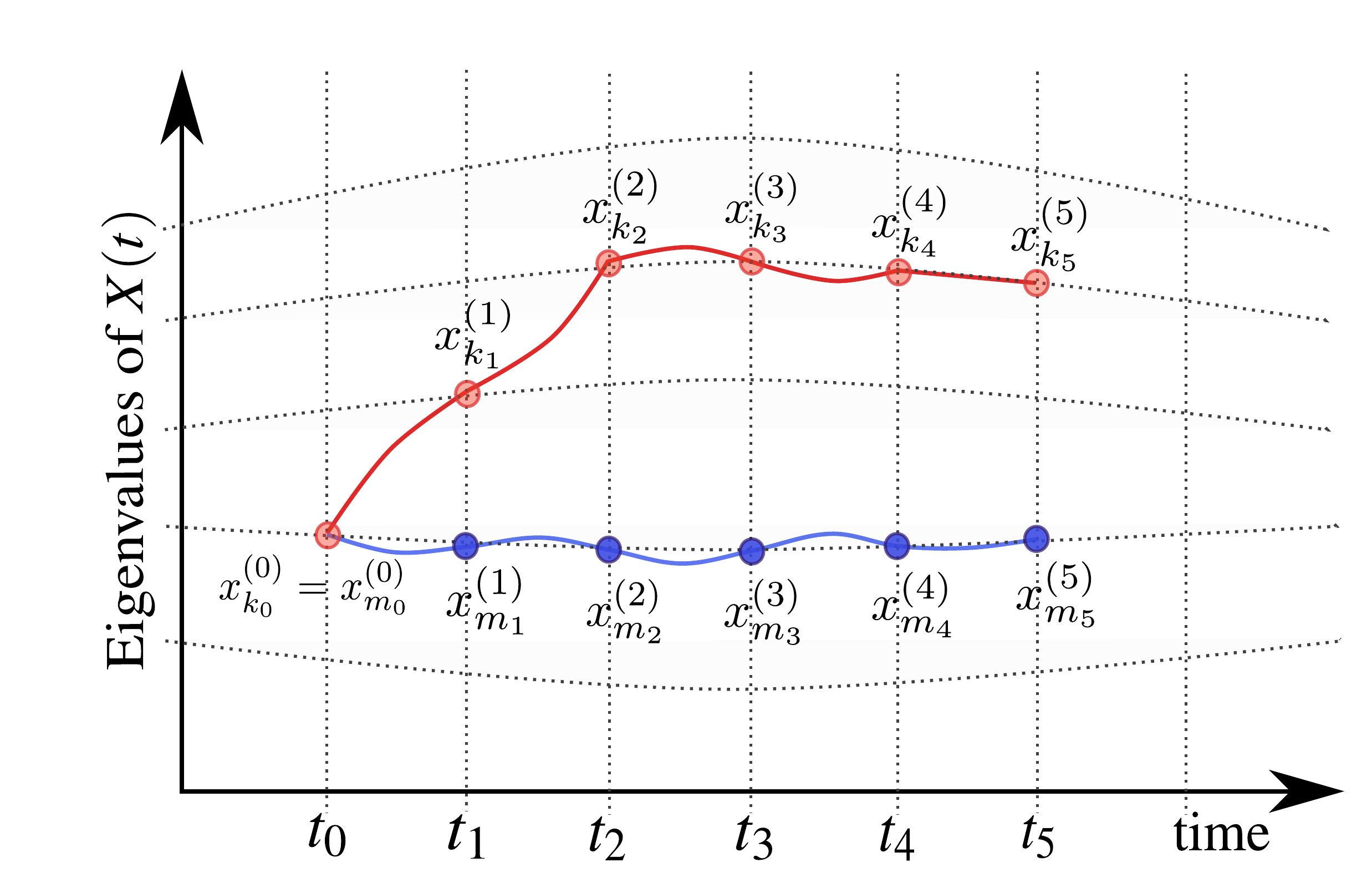}
\caption{\label{fig:histories} Trajectories $\vec{m}=\lbrace x^{(0)}_{m_{0}},x^{(1)}_{m_{1}},...,x^{(5)}_{m_{5}} \rbrace$ (blue) and $\vec{k}=\lbrace x^{(0)}_{k_{0}},x^{(1)}_{k_{1}},..,x^{(5)}_{k_{5}} \rbrace$ (red) denote sequences of eigenvalues of the operators $X_H\uj$ for a discrete set of times $t=\lbrace t_{0},t_{1},...,t_{5} \rbrace$. For example, trajectory $\vec{m}$ stays in the same eigenstate of the observables $X_H\uj$ at each step in time while trajectory $\vec{k}$ transfers between eigenstates.}
\end{figure}

A \emph{trajectory} for an observable $X_H (t)$ is defined by a class operator \cite{HALLIWELL1995}
\be \label{class}
	C_{\vec{n}}&=P_{n_{K}}\uk \, P_{n_{K-1}}\uko \, \cdot \, ... \cdot \, P_{n_{0}}\uz 
	= \overleftarrow{\mathcal{T}} \, \prod_{j=0}^{K-1} P_{n_{j}}\uj \, ,
\ee
where $\vec{n} = (n_0, n_1, ..., n_K)$ denotes the indices of a sequence of eigenvalues through which the trajectory passes at times $t_j$, see Fig.~\ref{fig:histories}. Here  $\overleftarrow{\mathcal{T}}$ is the time-ordering operator that arranges the product of projectors from right to left with increasing time. Trajectories are distinct when at least one eigenvalue $n_j$ in their sequence differs. The set of all distinct trajectories, $\{\vec{n}\}$, forms a complete set, i.e.
\be \label{complete}
	\sum_{\vec{n}} C_{\vec{n}} 
	= \sum_{n_0} ... \sum_{n_K} \, P_{n_{K}}\uk \, P_{n_{K-1}}\uko \, \cdot \, ... \cdot \, P_{n_{0}}\uz 
	= \mathbbm{1},
\ee
despite the class operators $C_{\vec{n}}$ not themselves being projection operators. Following Goldstein and Page \cite{Goldstein1995} we now associate a distribution for the trajectories similar to probability distributions  associated with classical trajectories:
\be\label{linear}
	p_{\vec{n}} : = {1 \over 2} \trace[ (C_{\vec{n}}^{\dag} + C_{\vec{n}}) \, \rho] \, ,
\ee
where $\rho$ is the initial state and the dynamics is captured by the class operators $C_{\vec{n}}$. This distribution is real and normalised, $\sum_{\vec{n}} p_{\vec{n}} = 1$, due to the completeness of the class operators. However, it has been shown that for certain choices of initial state $\rho$ and Hamiltonian $H (t)$ not all trajectories can be assigned a \emph{positive} probability \cite{Hartle2004,Hartle2008}. Therefore the distribution $p_{\vec{n}}$ is in fact a quasi-probability distribution sharing many features with the quasi-probability distributions associated with quantum states in quantum optics \cite{Lee1995}. A demonstration of the negativity of $p_{\vec{n}}$ will be shown later in Section \ref{sec:example}. 

It is now possible to group trajectories into \emph{histories} by defining a class operator, $C_{\alpha}$, for any set, $\alpha$, of distinct trajectories $\{\vec{n}\}$ as the sum over the class operators of the trajectories in that set \cite{HALLIWELL1995}, 
\be \label{cg}
	C_{\alpha}=\sum_{\vec{n} \in \alpha} \, C_{\vec{n}} \, . 
\ee 
Due to the linearity of Eq.~(\ref{linear}), the probability for any set of trajectories is then given by the sum of the individual probabilities for each trajectory in the set,
\be \label{eq:add}
	p_{\alpha} = \sum_{\vec{n} \in \alpha} p_{\vec{n}}.
\ee
By construction $p_{\alpha}$ is real and bounded from above by $1$ \cite{Hartle2004}, but can have negative values.

\section{A quantum work distribution from power trajectories}\label{sec:pow}

We are now ready to use the histories approach to identify a quantum work distribution for the closed quantum dynamics, i.e. the work that is done on the system when no measurements are made. 

The classical notion of work as power integrated along a particular trajectory in time motivates the choice of the power operator, $X(t) := \dot{H} (t)$, as the observable to construct histories with  \cite{Solinas2013,Li2013,PrasannaVenkatesh2015a}. Adopting the notation of the previous section we discretise the power operator in time and change into Heisenberg picture, $X_H\uj =  {V\uj}^{\dag} \, \dot{H}\uj \, V\uj$, where $\dot{H}\uj ={\partial H\uj \over \partial t}$. At each time $t_j$ the power operator eigenvalues are denoted $x_{n_j}\uj$ with corresponding projectors $P_{n_j}\uj$ with $n_j = (1, ..., d)$ labelling the eigenvalues, see Fig.~\ref{fig:histories}. The \emph{fluctuating work} $w_{\vec{n}}$ for a particular power trajectory specified by $\vec{n} = (n_0, n_1, ..., n_K)$, is now identified as the eigenvalue of the power operator at each time step multiplied by the length of the time step $\Delta t$,  summed over all time steps,
\be  \label{eq:flucw}
	w_{\vec{n}}=\sum_{j=0}^{K-1} x_{n_{j}}\uj \, \Delta t.
\ee

Following Eq.~(\ref{cg}) we now define the class operator for a {\it work history} as
\be\label{classw}
	C_{w}=\sum_{\vec{n}} \, \delta[w-w_{\vec{n}}] \, C_{\vec{n}} \, ,
\ee
where the sum is over the class operators  $C_{\vec{n}}$ of all power trajectories  that have the same work value $w_{\vec{n}} = w$. We finally obtain the  distribution of the quantum fluctuating work (\ref{eq:flucw}) based on multi time-step power trajectories (\ref{class}),
\be\label{workdist}
	p(w)
	&=&\frac{1}{2}\trace[(C_{w}^{\dag} + C_{w}) \, \rho] \\
	&=& \Re \left( \sum_{\vec{n}} \delta[w - w_{\vec{n}}] \, \trace[P_{n_{K}}\uk \, P_{n_{K-1}}\uko \, \cdot \, ... \cdot \, P_{n_{0}}\uz \, \rho] \right). \nonumber
\ee
This work distribution for the unmeasured dynamics will be central for the subsequent discussion. It allows the calculation of the average work and higher moments according to  Eq.~(\ref{eq:avgs}), as well as the discussion of properties of work under time-reversal.

\medskip

{\bf Energy conservation.} The average work obtained with $p(w)$ becomes
\be \label{eq:avgwork}
	\langle w \rangle 
	&=& \Re \left( \sum_{\vec{n}} w_{\vec{n}} \, \trace[P_{n_{K}}\uk \, P_{n_{K-1}}\uko \, \cdot \, ... \cdot \, P_{n_{0}}\uz \, \rho] \right) \nonumber \\
	&=& \Delta t \,  \Re \left( \sum_{j=0}^{K-1}  \,  \sum_{\vec{n}} \, \trace[P_{n_{K}}\uk \, \cdot \, ... \, \cdot \, x_{n_{j}}\uj \, P_{n_{j}}\uj \, \cdot \,  ... \cdot \, P_{n_{0}}\uz \, \rho] \right)  \nonumber \\
	&=& \Delta t \,   \sum_{j=0}^{K-1}  \,  \trace[X_H\uj \, \rho],
\ee
where we have used the identities $\sum_{n_j} P_{n_{j}}\uj = \mathbbm{1}$ and $\sum_{n_j} x_{n_{j}}\uj\, P_{n_{j}}\uj = X_H\uj$, and that the power operator is Hermitian. 

In the limit of infinitesimally small time steps, $\Delta t \to 0$, for a fixed time interval $\tau$ we replace $\Delta t \,  X_H\uj \to (H_H\ujp -H_H\uj)$ and obtain 
\be 
	\langle w \rangle = \sum_{j=0}^{K-1}  \,  \trace[(H_H\ujp -H_H\uj) \, \rho]
	= \trace[(H_H(\tau)-H_H(0)) \, \rho], \nonumber
\ee
where $H_H\uk = H_H(\tau)$ and $H_H\uz =H_H(0)$. We note that it is the linearity of the probabilities in the class operators, see Eq.~(\ref{linear}), that allows the explicit summation over the trajectories and this results in the contraction of the sum over power operators to the initial and final Hamiltonian only. This confirms that the work distribution (\ref{workdist}) in the continuum limit $\Delta t \to 0$ reproduces the desired average work,  Eq.~(\ref{1law}),  independent of the choice of state and unitary process.

\medskip

{\bf Time-reversal symmetry.}
We now address the second desirable property for any definition of work, namely time-reversal symmetry (ii). Physically ``time-reversal'' refers to a reversal of the order of events in the Heisenberg picture. In the histories approach the work history of a time-reversed closed process, determined by the discretised sequence of power operators $X_H\uk, X_H\uko, ..., X_H\uz$, can naturally be assigned the class operator \cite{Goldstein1995}
\be
	C^{\dag}_{-w} = \sum_{\vec{n}} \, \delta[w + w_{\vec{n}}] \, P_{n_{0}}\uz \, P_{n_{1}}\uo \, \cdot \, ... \cdot \, P_{n_{K}}\uk.
\ee
Here the projectors are time-ordered in reverse order, i.e. the time-reversed power trajectories are $C_{\vec{n}}^{\dag} =  P_{n_{0}}\uz \, P_{n_{1}}\uo \, \cdot \, ... \cdot \, P_{n_{K}}\uk$. Also, the work value associated with each time-reversed trajectory has obtained a minus sign, see Eq.~(\ref{eq:flucw}), as time has inverted, $\Delta t \to -\Delta t$. 

Applying this time-reversal to the work distribution Eq.~(\ref{workdist}) we trace the class operator with the same state $\rho$ and rearrange to obtain the work distribution of the unmeasured time-reversed process 
\be\label{eq:reverse}
	p_{\back}(w) = \frac{1}{2}\trace[(C_{-w}  + C^{\dag}_{-w} ) \, \rho] = p(-w).
\ee
This shows that in the histories approach the time-reversed process has a work distribution that is equal to that of the forward process with a negative work value \cite{Goldstein1995, Gell-Mann2012}. This relation reflects the fundamental time symmetry of the statistics of work in a closed quantum system undergoing unitary dynamics.

\medskip

{\bf Positivity.} 
As shown in \cite{Hartle2004} the Hermitian part of a product of projectors, cf. (\ref{linear}), generally has at least one negative eigenvalue and that can lead to negative probabilities, $p_{\vec{n}}$, for some power-trajectories. While the summation over various power trajectories may result in a positive work distribution $p(w)$ there are examples of closed quantum dynamics where $p(w)$ is negative for some work values thus making it a quasi-probability distribution.  
An example of the negativity of the work distribution is discussed in Section \ref{sec:example}. 

The occurrence of quasi-probability distributions used to represent a quantum {\it state} in phase space, such as the Wigner and Dirac distributions, is a typical quantum signature \cite{Lee1995,Garraway1993}. These quasi-probabilities contain physical information about the state of the system despite not referring directly to measurement outcomes. For example, the Wigner function cannot be observed from direct measurements of position and momentum. Instead parity measurements are made to reconstruct it experimentally \cite{Banaszek1999}. Furthermore quasi-probabilities naturally emerge from weakly measured quantum systems. For example, the Dirac function has been determined for photon states by weak measurements of both position and momentum \cite{Lundeen2012,Bamber2014}.  

In the present context the emergence of a quasi-probability distribution characterises non-classical behaviour of a non-equilibrium {\it process} rather than a state. The presence of negative ``probabilities'' here indicates that, in general, quantum fluctuating work cannot correspond to measurement outcomes, i.e. there is no POVM whose outcomes refer to fluctuating work values. Thus work is an indeterminate property of a quantum process, in contrast to classical notions of work. These findings are in agreement with a recent no-go proof by Perarnau-Llobet \textit{et al.} showing that it is not possible to assign a POVM to fluctuating work in an isolated quantum system whilst maintaining average energy conservation for all initial states alongside consistency with classical stochastic thermodynamics \cite{Perarnau-Llobet2016}.

\medskip

{\bf Higher moments of work.}
The work distribution Eq.~(\ref{workdist}) has the desired properties (i) and (ii) given in the Introduction. We now discuss further statistical properties of the work distribution, such as higher moments and average exponentiated work, in the following. To determine the spread of the work distribution requires the second moment of work, given as 
\be
	\langle w^{2} \rangle 
	&=& \Re \left( \sum_{\vec{n}} w^2_{\vec{n}} \, \trace[P_{n_{K}}\uk \, P_{n_{K-1}}\uko \, \cdot \, ... \cdot \, P_{n_{0}}\uz \, \rho] \right) \nonumber \\
	&=& (\Delta t)^2 \,  \Re \left( \sum_{j,i=0}^{K-1}  \,  \sum_{\vec{n}} \, x_{n_{j}}\uj  \, x_{n_{i}}\ui  \trace[P_{n_{K}}\uk  \, P_{n_{K-1}}\uko \, \cdot \, ... \cdot \, P_{n_{0}}\uz \, \rho]  \right)  \nonumber \\
	&=& (\Delta t)^2 \, \sum_{j,i=0}^{K-1} \trace[X_H\ui \, X_H\uj \, \rho].
\ee
In the last line we have used that the expression is real as the two sums run separately, $\sum_{j=0}^{K-1} X_H\uj$,  and contract to the same Hermitian operator. In the continuum limit $\Delta t \to 0$ the second moment simplifies to $\langle w^{2} \rangle = \trace[(H_H (\tau)-H_H (0))^{2} \, \rho]$.  It is straightforward to generalise this derivation to the $m$-th moment of work for $m = \lbrace 1, 2, ... \rbrace$
\be   \label{moments}
	\langle w^{m} \rangle=\trace[(H_H(\tau)-H_H(0))^{m} \, \rho],
\ee
in the continuum limit $\Delta t \to 0$.

Eq.~(\ref{moments}) implies that for each $m$ the value of $\langle w^{m} \rangle$ can be measured as the expectation value of the observable $\big(H_H(\tau)-H_H(0)\big)^m$ for state $\rho$. The quantum work distribution can then be inferred  from the moment-generating function $G(\lambda)$ \cite{Esposito2009},
\be
	p(w) = \int G(\lambda) \, e^{-i\lambda w} \, \d \lambda, 
\ee
where $G(\lambda)$ can be constructed from the measured $\langle w^{m} \rangle$ as
\be\label{momg}
	\nonumber G(\lambda)&:=&\langle e^{i\lambda w} \rangle 
	= \sum^{\infty}_{m=0}\frac{(i\lambda)^{m}}{m!}\langle w^{m} \rangle \\
	&=&\trace[e^{i\lambda(H_H(\tau)-H_H(0))} \, \rho ],
\ee
where the summation has been carried out in the last line. 

From expressions (\ref{moments}) and (\ref{momg}) one may be tempted to define fluctuating work values as the eigenvalues of a ``work operator'' $W = \sum_k w_k \, P_k=H_H(\tau)-H_H(0)$, where $P_k$ are a set of projectors onto ``work'' eigenstates, as discussed in \cite{Allahverdyan2005a,Allahverdyan2014c}. As a consequence, in the work operator approach the fluctuating work values would be the eigenvalues  $w_k$ of $W$. In this case work is identified as a property of the state rather than the process applied to the state, with the range of possible work values given by the dimension of the Hilbert space, in contrast to the TPM scheme \cite{Talkner2007c}. This runs counter to the usual assumption that thermodynamic work is a process-dependent quantity, and so the work operator is often considered as a conceptually less favourable definition of quantum work \cite{Allahverdyan2005a,Allahverdyan2014c}. In contrast to the work operator, the histories framework explicitly defines work in terms of process-dependent trajectories that are not properties of the state alone, and the range of the possible work values $w_{\vec{n}}$ greatly outweighs the dimension of the Hilbert space. The benefit of this approach is that it is more closely tied to the classical thermodynamic definition of work as power integrated along a given process. It is interesting to observe that while both the work operator and histories approach produce the same set of work moments, the statistical distributions are of course different. In particular, projective measurements of the work operator give a positive-definite probability distribution, in contrast to the quasi-probability Eq.~(\ref{workdist}). Thus the two approaches significantly differ at the level of the individual probabilities assigned to each value of fluctuating work.

\medskip

Because of its central importance for the study of non-equilibrium work in the classical regime, it is of interest to see whether the Jarzynski equality is satisfied by the quantum work distribution $p(w)$ defined in Eq.~(\ref{workdist}).  In the standard Jarzynski setup \cite{Jarzynski1997d,Talkner2007c}, the initial state is chosen to be thermal with respect to an environment at inverse temperature $\beta$, i.e.  $\rho= \trace[e^{-\beta H(0)}]/Z_0$ with $Z_{0}=\trace[e^{-\beta H(0)}]$ the system's partition function.  For a closed system that is brought out of equilibrium by its initial Hamiltonian becoming time-dependent, $H(t)$, the classical Jarzynski equality relates the average exponentiated work done on the system with an equilibrium free energy difference, $\langle e^{- \beta w} \rangle=e^{- \beta \Delta F}$. Here $\Delta F= {1 \over \beta}  \ln (Z_{0} / Z_{\tau})$ is the free energy change associated with the initial thermal state and a thermal state defined for the final Hamiltonian, $H(\tau)$, at the same inverse temperature  $\beta$ with partition function $Z_{\tau}=\trace[e^{-\beta H(\tau)}]$. 

Using the work distribution $p(w)$ derived with the histories approach and taking the continuum limit $\Delta t \to 0$, the average exponentiated work can be obtained by setting $\lambda=i\beta$ in Eq.~(\ref{momg}),
\be\label{expw}
	\langle e^{- \beta w} \rangle=\trace[e^{-\beta(H_H(\tau)-H_H(0))} \, \rho].	
\ee
Inserting a thermal initial state one can see that the standard Jarzynski equality is no longer generally satisfied,
\be	\label{jarz}
	\langle e^{- \beta w} \rangle 
	&=& \trace[e^{-\beta(H_H(\tau)-H_H(0))} \, e^{-\beta H(0)}]/ Z_0  \nonumber\\ 
	&\geq& \trace[e^{-\beta H_H(\tau)}] / Z_0 =e^{- \beta \Delta F},
\ee
where $H_H  (0)= H(0)$ and the identity $\trace[e^{-\beta H_H(\tau)}] = \trace[V^{\dag} ({\tau}) \, e^{-\beta H(\tau)} \, V({\tau})] = Z_{\tau}$ has been used. Here the Golden-Thompson inequality $\trace [e^{A}e^{B}]\geq \trace [e^{A+B}]$ has been applied where equality holds if and only if $A$ and $B$ commute \cite{Cohen1982}. Thus, in general, the average exponentiated work is \emph{greater} than the exponentiated free energy change. The standard Jarzynski equality for an initial thermal state only holds when the Hamiltonians (in Heisenberg picture) commute, $[H_H(\tau),H_H(0)]=0$. In  Schr\"odinger picture this condition translates into $[H (\tau), V \, e^{-\beta H(0)} \, V^{\dag} ]=0$, i.e. the final state after the unitary evolution must be diagonal for the final Hamiltonian \cite{Kammerlander2016}. This condition can be thought of as the classical limit demonstrating that one recovers the original Jarzynski equality.

Inequality (\ref{jarz}) based on the work distribution $p(w)$ thus exposes quantum signatures of work in general quantum non-equilibrium processes. This contrasts with the two-point measurement approach, in Eq.~(\ref{eq:tpmw}), which shows the validity of the (classical) Jarzynski equality for an initial thermal state and any choice of time-dependent Hamiltonians \cite{Talkner2007c}. This result demonstrates that when extending fluctuation relations to the quantum regime modifications can arise from coherences, as has been suggested by Aberg \cite{Aberg2016}. We note that the derivation of inequality (\ref{jarz}) assumes an initial thermal state, contrasting with previous discussions of quantum signatures in the average exponentiated work by Allhaverdyan, Solinas {\it et al.} and Elouard {\it et al.} that have focussed on non-thermal initial states \cite{Allahverdyan2014c,Solinas2015,Elouard2015}. The properties of $p(w)$ are listed in Table~\ref{tab:compare} and will be compared with another power-operator-based work distribution \cite{Li2013,PrasannaVenkatesh2015a} in Section~\ref{sec:comparison}. Before that we will first discuss the connection to the TPM definition of work.

\medskip

{\bf Recovering the TPM work distribution.} 
Here we illustrate for a particular example the underlying difference between the two distributions, the measured $p^{\tpm}(w)$ and the unmeasured $p(w)$. We let the Hamiltonian depend on time solely through its eigenvalues while the projectors on the eigenstates are time-invariant, i.e. $H(t) = \sum_n E_n(t) \, \Pi_n$. The unitary generated by this Hamiltonian is then $V(t) = e^{- i \sum_n \phi_n(t) \, \Pi_n}$. The power operator (in either Heisenberg or Schr\"odinger picture) is then $X(t) = \sum_n \dot{E}_n(t) \, \Pi_n$ and the time-discretised eigenvalues belonging to the projectors $\Pi_{n_j}\uj$ are defined as $x_{n_j}\uj = \dot{E}_{n_j} (t_j) = (E_{n_j}\ujp - E_{n_j}\uj)/\Delta t$ in the limit $\Delta t \to 0$. These power operator eigenvalues describe the variation of a specific energy eigenvalue, labelled with $n_j$, with time $j \to j+1$. Using the power operator definition of the fluctuating work, Eq.~(\ref{eq:flucw}), for a trajectory $\vec{n}$ one obtains 
\be  
	w_{\vec{n}}
	&=& \sum_{j=0}^{K-1} (E_{n_j}\ujp - E_{n_j}\uj) \nonumber \\
	&=& \sum_{j=0}^{K-1} (E_{n_{j+1}}\ujp - E_{n_j}\uj) - \sum_{j=0}^{K-1} (E_{n_{j+1}}\ujp - E_{n_j}\ujp)\\
		&=:& E_{n_K}\uk - E_{n_0}\uz - \Delta E_{\vec{n}}. \nonumber
\ee
In the second line we have split the sum into two terms. The first term describes the energetic jumps from level $n_j$ at time $j$ to a different level, i.e. $n_{j+1}$, at time $j+1$. The second term, for which the symbol $\Delta E_{\vec{n}}$ has been introduced, describes the energetic difference between the two levels $n_j$ and $n_{j+1}$ belonging to different projectors $\Pi_{n_j}\ujp$ and $\Pi_{n_{j+1}}\ujp$ of a single Hamiltonian at time step $j+1$. 

While the second term vanishes when summed over all trajectories,
\be  
	\sum_{\vec{n}} \Delta E_{\vec{n}} 
	\propto \sum_{j=0}^{K-1} \left(\sum_{n_{j+1}}  E_{n_{j+1}}\ujp - \sum_{n_{j}} E_{n_j}\ujp \right)
	= 0,
\ee
the work distribution (\ref{workdist}) will depend on individual $\Delta E_{\vec{n}}$. Only when this dependence is neglected and under an additional assumption (one pair of the three operators $\rho, H_0$ and $H_H(\tau)$ have to commute) can the TPM work distribution be recovered, as will be discussed in Section \ref{sec:MH}.

\section{Comparison with other power-operator-based work distributions}\label{sec:comparison}

The histories framework and the resulting work distribution $p(w)$ for the unmeasured dynamics differ from previous proposals to define fluctuating work using the power operator \cite{Chernyak2004,Subas2012,Li2013,PrasannaVenkatesh2015a}. In those proposals the power is continuously {\it measured} leading to a work probability distribution given as
\be \label{eq:tildep}
	\tilde{p}(w) =\sum_{\vec{n}} \, \delta \left[w- \sum_{j=0}^{K-1} x_{n_{j}}\uj \, \Delta t \right] \, \trace[C^{\dag}_{\vec{n}} \, C_{\vec{n}} \, \rho].
\ee
This distribution differs from $p(w)$ in Eq.~(\ref{workdist}) by the probability $\tilde{p}_{\vec{n}} = \trace[C^{\dag}_{\vec{n}} \, C_{\vec{n}} \, \rho]$ which is \emph{non-linear} in the class operators $C_{\vec{n}}$ assigned to each power trajectory $\vec{n}$. 
As we will see in this section the distribution $\tilde{p}(w)$ is normalised and also positive, but  $\tilde{p}(w)$ does not satisfy the desired physical conditions (i) and (ii).

\medskip

{\bf Energy conservation.}
We illustrate that the work probability distribution $\tilde{p}(w)$ does not generally satisfy average energy conservation, Eq.~(\ref{1law}), using an example. 
Let the Hamiltonian of the system be of the form $H(t)=A+\lambda (t) \, B$, where $A$ and $B$ are time-independent Hermitian operators while $\lambda (t)$ is a real, time-dependent scalar function \cite{PrasannaVenkatesh2015a}, then the power operator in the Heisenberg picture is $X_{H}(t)=\dot{\lambda}(t) \, B_{H}(t)$. At each discrete point in time, labelled again by $j$, the operator $B_{H}\uj=\sum_{n_{j}} b_{n_{j}}\uj \, P_{n_{j}}\uj$ is spectrally decomposed into its eigenvalues $b_{n_{j}}\uj$ and projectors $P_{n_{j}}\uj$. As shown in \cite{PrasannaVenkatesh2015a} in the limit of small $\Delta t$ the work distribution $\tilde{p}(w)$ becomes
\be  \label{eq:tp0}
	\tilde{p}(w) = \sum_{n_{0}}\delta[w-(\lambda(\tau)-\lambda(0)) \, b_{n_{0}}\uz \, ] \, \trace[ P_{n_{0}}\uz \, \rho].
\ee
Here only the eigenvalues $b_{n_{0}}\uz$ of the power operator $B_H\uz = B$ at the start of the evolution appear. Using the probability distribution $\tilde{p}(w)$ the average work done on the system is,
\be \label{eq:avgwtilde}
	\langle w\rangle_{\tilde{p}(w)}
	&=& \trace[\lambda(\tau) B \, \rho] - \trace[\lambda(0) B \, \rho] \nonumber \\
	&=& \trace[H (\tau) \, \rho] - \trace[H (0) \, \rho].
\ee
This average work differs in general from the first law stated in Eq.~(\ref{1law}) because the final Hamiltonian, $H (\tau)$, is here in the Schr\"odinger picture but the expectation value is taken for the initial state rather than for the time-evolved state. Indeed as discussed in \cite{PrasannaVenkatesh2015a}, the physical situation characterised by the distribution $\tilde{p} (w)$ is one where the dynamics of the system is frozen by the initial projective measurement. The occurrence of this quantum Zeno effect is reflected in the expressions in Eqs.~(\ref{eq:tp0}) and (\ref{eq:avgwtilde}).  
It can be seen that $\tilde{p}(w)$ will satisfy energy conservation for all initial states $\rho$ in the special case that $H_{H}(\tau)=H(\tau)$. Like $p(w)$ the distribution $\tilde{p}(w)$ is found \cite{PrasannaVenkatesh2015a} not to obey the Jarzynski equality unless one has $H_{H}(\tau)=H(\tau)$. 

\medskip

{\bf Time-reversal symmetry.}  
The work distribution for the time-reversed process, $\tilde{p}_{\back}(w)$, is defined by exchanging each class operator $C_{\vec{n}}$ in Eq.~(\ref{eq:tildep}) with its time-reversed counterpart $C_{\vec{n}}^{\dag}$, see Eq.~(\ref{class}), and changing the sign for $w$ to reflect the fact that we are considering the time reversed unitary evolution. 
One obtains 
\be
	\tilde{p}_{\back}(w) 
	&=& \sum_{\vec{n}} \, \delta \left[w+ w_{\vec{n}} \right] \, \trace[C_{\vec{n}} \,C_{\vec{n}}^{\dag} \, \rho] \\
	&=& \tilde{p}(-w)+\sum_{\vec{n}} \, \delta \left[w+w_{\vec{n}} \right] \, \trace[[C_{\vec{n}}, C_{\vec{n}}^{\dag}] \, \rho], \nonumber
\ee
where the work $w_{\vec{n}}$ for each power trajectory is again given by Eq.~(\ref{eq:flucw}), cf. (\ref{eq:tildep}). For the work distribution Eq.~(\ref{eq:tildep}) to satisfy time-reversal symmetry, the class operators for the power histories must obey $\trace[ \sum_{\vec{n} \in w} \, [C_{\vec{n}}, C_{\vec{n}}^{\dag}] \, \rho] = 0$ for all work values $w$. We will discuss an example showing that this condition is in general not fulfilled in the next section. The physical properties of $p(w)$ and $\tilde{p}(w)$ are listed in Table~\ref{tab:compare}. 

\begin{table*}[t]
\centering
\captionsetup{margin=1cm}
\scalebox{0.75}{
\begin{tabular}{
| >{\hspace{\ssp}} l <{\hspace{\ssp}}
| >{\hspace{\ssp}} l <{\hspace{\ssp}} 
| >{\hspace{\ssp}} l <{\hspace{\ssp}} 
| >{\hspace{\ssp}} l <{\hspace{\ssp}} 
| >{\hspace{\ssp}} l <{\hspace{\ssp}} 
| >{\hspace{\ssp}} l <{\hspace{\ssp}} 
| >{\hspace{\ssp}} l <{\hspace{\ssp}} 
|} 
\hline
i. 
& ii. 
& iii. operator
& iv. steps
& v. energy conservation 
& vi. time-reversal symmetric
& vii. JE satisfied  
 \\
\hline
$p^{\tpm} (w)$ & +   & $H$ & $2$   & $\forall \rho$, $ [H_{H}(\tau), H(0)]=0$ & $\forall \rho$, $ [H_{H}(\tau), H(0)]=0$  & $\forall H(t)$  \\ 
 & & & & or $\forall H_{H}(\tau)$, $[\rho, H(0)]=0$ & or $\forall H_{H}(\tau)$, $[\rho, H(0)]=0$ & \\ \hline
$p(w)$ & - & $X$ & $K \to \infty$  & $\forall \rho$, $\forall H(t)$ & $\forall \rho$, $\forall H(t)$ & $[H_{H}(\tau), H(0)]=0$  \\ \hline
$\tilde{p}(w)$  & + & $X$ & 1   &  $\forall \rho$, $H_{H}(\tau)=H(\tau)$  & $\trace[\sum_{\vec{n} \in w} \,  [C_{\vec{n}}, C_{\vec{n}}^{\dag}] \, \rho] =0$, $\forall w$&  $H_{H}(\tau)=H(\tau)$\\ \hline
 $p^{\MH} (w)$  & -  & $H$ & $2$      & $\forall \rho$, $\forall H(t)$ & $\forall \rho$, $\forall H(t)$   & $\forall H(t)$    \\ \hline
\end{tabular}
}
\hspace*{-4cm}
\caption{\label{tab:compare} 
Comparison between the work \emph{quasi-probability distributions} $p(w)$ given in Eq.~(\ref{workdist}) and $p^{\MH}(w)$ given in Eq.~(\ref{mh}) and the work probability distributions $p^{\tpm}(w)$ given in Eq.~(\ref{eq:tpmw}) and $\tilde{p}(w)$ given in Eq.~(\ref{eq:tildep}). The columns refer to: (i) symbol of work distribution, (ii) work distribution is  positive in general (+) or can be negative for some states and evolutions (-), (iii) Operator used to establish the work distribution: power operator $X$ or Hamiltonian $H$, (iv) number of points in time considered in the construction of the work distribution for $\Delta t \to 0$. Columns (v), (vi) and (vii) list conditions under which the work distributions satisfy energy conservation in the limit $\Delta t \to 0$, time-reversal symmetry, and the Jarzynski equality (JE) assuming an initial thermal state and the limit $\Delta t \to 0$. 
All four work distributions become equal in the classical limit where $\rho, H_{H}(0)$ and $H_{H}(\tau)$ commute.}
\end{table*}

\section{Example: Two-level system in an oscillating field} \label{sec:example}

To illustrate the histories approach and the negativity of the many-point work distribution $p (w)$ we here discuss the example of a qubit driven by an oscillating field,
\be \label{ham}
	H(t) = \frac{\Omega}{2}\sigma_{z} + \frac{g}{2} (\cos (\Omega \, t) \, \sigma_{x} + \sin (\Omega \, t)\, \sigma_{y} ),
\ee
where $\sigma_{x,y,z}$ are the Pauli spin matrices, $\Omega$ is the free frequency of the qubit system and $g$ is the coupling strength of the qubit to the external field. Allahverdyan previously used this model \cite{Allahverdyan2014c} to demonstrate the negativity of a two-point work distribution denoted by $p^{\MH}(w)$, which is defined below in Eq.~(\ref{mh}). Here we will consider a discretised form of the work distribution Eq.~(\ref{workdist}) in order to demonstrate the negativity of the distribution. The time-dependent Hamiltonian generates a unitary 
\be
	V(t)=e^{ -i \, \Omega \, t \, \sigma_{z}/ 2} \, e^{ -i \, g \, t \, \sigma_{x}/2},
\ee
which can be verified by substituting into the evolution equation, $i \, \partial_{t}V(t)= H(t) \, V(t)$ \cite{Allahverdyan2014c}. 
The power operator in the Heisenberg picture is then given by
\be
	X_{H}(t) 
	&=&\frac{g\Omega}{2} (\cos (gt) \, \sigma_{y} -\sin (gt) \, \sigma_{z}),
\ee
and has a time-dependent eigenbasis while the eigenvalues, $x =\pm \, g\Omega / 2$, are time-independent. 

To simplify the subsequent calculation we choose to discretise time into time steps of length $\Delta t=\pi/(2g)$, such that the time-discretised power operators oscillate between $\sigma_y$ and $\sigma_z$. I.e. for the first four time steps the power operator becomes $X_H\uz= (g\Omega/2) \, \sigma_{y}$, $X_H^{(1)} = - (g\Omega/2) \, \sigma_{z}$, $X_H^{(2)}= - (g\Omega/2) \, \sigma_{y}$ and $X_H^{(3)} = (g\Omega/2) \, \sigma_{z}$ and then this sequence repeats every set of four steps. The eigenstates of $X_H\uj$ with eigenvalues $(-1)^{n_j} g\Omega / 2$, for $n_j = 0, 1$, can be cast in the form
\be
\ket{j,n_{j}} &=&
	\sum^{K-1}_{l=0}  (-1)^{l} \,  \left(
			\delta_{j,2l} \, \ket{(-1)^{n_j} i \, } - \delta_{j,2l+1} \, \ket{n_{j}}
		\right).
\ee
Depending on $j$ being even or odd, the eigenvectors will be either the eigenvectors $\ket{+i} = (\ket{0} + i \ket{1} )/ \sqrt{2}, \ket{-i} = (\ket{0} - i \ket{1})/ \sqrt{2}$ of $\sigma_y$  or the eigenvectors $\ket{0}, \ket{1}$ of $\sigma_z$ for the eigenvalues $+1, -1$. The total time of evolution, $\tau$, is determined by the total number of time steps, $K$, i.e. $\tau = K \, \pi/(2g)$.

\begin{figure}[t]
\hspace*{-1.3cm} 
\centering
\includegraphics[scale=0.5]{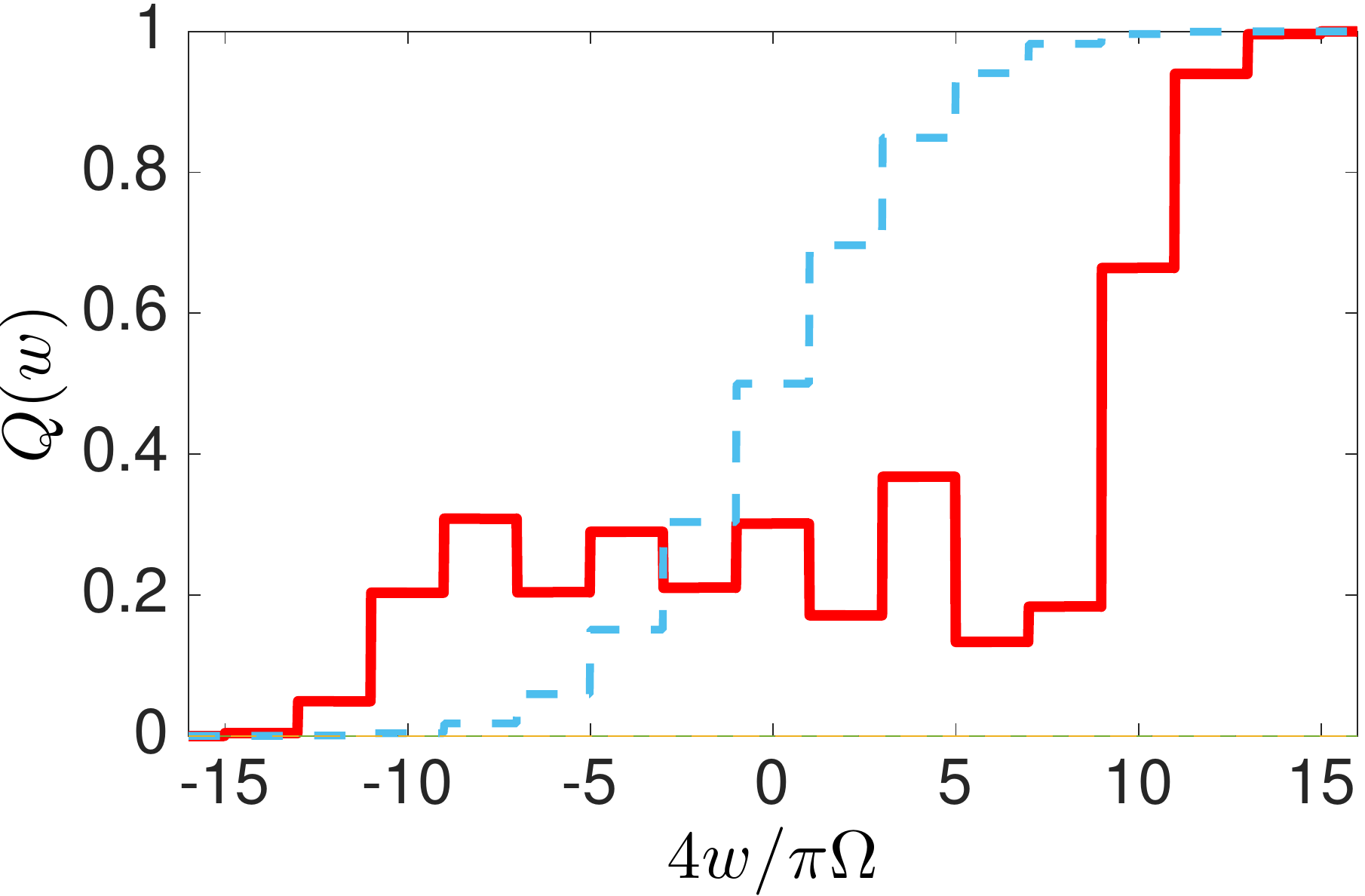}
\centering
\caption{\label{fig:cumulative} 
Plot of cumulative work distribution $Q(w)$, see text, over work values $w$ for the work distribution $p(w)$ (red) and the work probability distribution $\tilde{p} (w)$ (blue), given in Eq.~(\ref{p1}) and Eq.~(\ref{p2}), respectively. Here the Hamiltonian is given in (\ref{ham}), the time steps are of length $\Delta t=\pi/2g$, the total number of steps is $K=15$, and the initial state is thermal with respect to the initial Hamiltonian, ie. $\rho \propto \text{exp}(-\beta H(0))$ with $\beta=0.1$. }
\end{figure}

From Eq.~(\ref{eq:flucw}) for a given sequence $\vec{n} = (n_0, n_1, ..., n_K)$ the fluctuating work in this example is $w_{\vec{n}}= (\pi \Omega / 4) \, \sum_{j=0}^{K-1} (-1)^{n_{j}}$. The many-point work distribution based on the histories approach becomes, according to Eq.~(\ref{workdist}),  
\be\label{p1}
	p(w) 
	&=& \sum_{\vec{n}} \, \delta \left[w- (\pi \Omega / 4) \, \sum_{j=0}^{K-1} (-1)^{n_{j}} \right] \, p_{\vec{n}} \, , \\
	\mbox{with} && p_{\vec{n}}=\Re \left[\bra{0,n_{0}}\rho\ket{K,n_{K}}\prod^{K-1}_{j=0}\braket{j+1,n_{j+1}|j,n_{j}}  \right],	\nonumber
\ee
where we have used the cyclic property of the trace. Meanwhile the probability distribution $\tilde{p}(w)$ obtained with continuous measurement of the power operator, Eq.~(\ref{eq:tildep}), is given by 
\be\label{p2}
	\tilde{p}(w) 
	&=& \sum_{\vec{n}} \, \delta \left[w- (\pi \Omega / 4) \, \sum_{j=0}^{K-1} (-1)^{n_{j}} \right] \, \tilde{p}_{\vec{n}} \, , \\
	\mbox{with} && \tilde{p}_{\vec{n}} = \bra{0,n_{0}} \rho \ket{0,n_{0}} \prod^{K-1}_{j=0} \left|\braket{j+1,n_{j+1}|j,n_{j}} \right|^{2}.	\nonumber
\ee

We immediately see that the structure of $p_{\vec{n}}$ in (\ref{p1}) and $\tilde{p}_{\vec{n}}$ in (\ref{p2}) is general when $|j, n_j \rangle$ denotes the eigenstates of general power operators $X_H\uj$. Differences between $p(w)$ and $\tilde{p}(w)$ can now be identified: $\tilde{p}_{\vec{n}}$ is a product of transition probabilities, $\left|\braket{j+1,n_{j+1}|j,n_{j}}\right|^{2}$, and thus $\tilde{p} (w)$ becomes a real and positive probability distribution. In contrast the distribution $p(w)$ is formed by (the real part of) a product of transition amplitudes, $\braket{j+1,n_{j+1}|j,n_{j}}$, and while it is real and normalised, the distribution can have negative values for some initial states $\rho$. This feature is illustrated in Fig.~\ref{fig:cumulative}, where the cumulative distributions, $Q(w)=\int_{-\infty}^{w} p(w') \d w'$ are plotted for $p(w)$ and $\tilde{p}(w)$ for an initial thermal state $\rho \propto \text{exp}(-\beta H(0))$. One can see that the cumulative distribution for $p(w)$ \emph{decreases} for certain values of work $w$, demonstrating the negativity of this quasi-probability distribution. Note that the breaking of the time-reversal symmetry of the distribution $\tilde{p}(w)$ is clearly evidenced by the fact that the initial probability for state $|0,n_0\rangle$ appears at the start of the chain, see Eq.~(\ref{p2}), contrasting with the symmetric inclusion of both $|0,n_0\rangle$ and $|K,n_K\rangle$ for $p(w)$ seen in Eq.~(\ref{p1}). 

In the next section we show that rather than defining a work distribution by time-integrating over continuous power-trajectories it is also possible to define a \emph{time-reversal symmetric} work quasi-probability distribution, denoted $p^{\MH}(w)$, that is based on just a two-time energy trajectory. We will find that in the special case that the initial state $\rho$ has no coherences with respect to the initial Hamiltonian $H(0)$ one recovers the standard TPM distribution, $p^\tpm (w)$.

\section{Two-point energy Margenau-Hill work distribution} \label{sec:MH} 

We now adopt the TPM's fluctuating work definition Eq.~(\ref{eq:tpm}) that refers to energy eigenvalues rather than power operator eigenvalues. Here we will consider the relevant observable $X(t)=H(t)$ as the system's time-dependent Hamiltonian, with $H_H\uj = \sum_{n_{j}=1}^d \eps_{n_j}\uj \, \Pi_{n_j}\uj$ denoting the spectral decomposition of the Hamiltonian at time $t_j$ in the Heisenberg picture. Here $\Pi_{n_{j}}\uj$ are the projectors onto the energy eigenstates of the Hamiltonians $H_H\uj$. This motivates the construction of energy trajectories rather than power trajectories and the class operator for a work history with fluctuating work values $w$ can now be identified as
\be \label{cw}
	C^{\tpm}_{w} =\sum_{n_0, n_K} \, \delta[w-(\eps_{n_K}\uk-\eps_{n_0}\uz)] \, \Pi_{n_{K}}\uk \, \Pi_{n_{0}}\uz. 
\ee
The summation over the intermediate time-steps has been carried out because the work value in the TPM scheme only depends on the initial and final energy. The remaining summation over initial and final energy eigenvalues makes sure that different energetic transitions with the same fluctuating work value are included. 

Using definition (\ref{eq:add}) the work distribution function now becomes
\be  \label{mh}
	p^{\MH} (w) =  \sum_{n_0, n_K} \, \delta[w-(\eps_{n_K}\uk-\eps_{n_0}\uz)] \, p^{\MH}_{n_0, n_K} 
\ee
where $p^{\MH}_{n_0, n_K}: = {1 \over 2} \trace[ (\Pi_{n_{K}}\uk \, \Pi_{n_{0}}\uz +  \Pi_{n_{0}}\uz \, \Pi_{n_{K}}\uk) \, \rho]$ are joint probabilities known as the \emph{Margenau-Hill distribution} \cite{Margenau1961}. This distribution is a generalisation of the two-point joint probability $p^{\tpm}_{n_0, n_K}$ given in Eq.~(\ref{eq:jointtpm}), that is associated with the outcomes of two successive projective measurements in the TPM scheme \cite{Johansen2007}. $p^{\MH}_{n_0, n_K}$ gives the correct marginals of initial and final state and recovers the TPM work probability distribution, $p^{\tpm}_{n_0, n_K}$, when at least one pair of the operators $\rho$, $H(0)$ or $H_{H}(\tau)$ commute. 
The assumption that one of the commutators is zero is often made, including derivations of the quantum Jarzynski equality  \cite{Talkner2007c,Esposito2009,Jarzynski2015a} where the initial state commutes with the initial Hamiltonian. 
In general, the work distribution $p^{\MH} (w)$ is not positive-definite. For unitary dynamics of states that do have coherences in the initial Hamiltonian basis the work distribution $p^{\MH}(w)$ has previously been proposed by Allahverdyan \cite{Allahverdyan2014c}, and was shown to minimise the mean-squared uncertainty between the final and the initial Hamiltonian. Another two-point work quasi-probability distribution that obeys the first law and fulfils the Jarzynski equality has been proposed by Solinas and Gasparinetti \cite{Solinas2015} which is based on the full-counting statistics. 

The quasi-probability distribution $p^{\MH} (w)$ gives the work moments
\be 	\label{mhmoments}
	 \langle w^{m} \rangle_{p^{\MH}(w)} 
	&= & \sum_{n_0, n_K}  (\eps_{n_K}\uk - \eps_{n_0}\uz)^{m} \, p^{\MH}_{n_0, n_K}  \\
	&=& \frac{1}{2} \,  \sum_{l=0}^{m}  \binom{m}{l} \, \trace \left[ \left\{ (H_H(\tau))^l , (- H_H(0))^{m-l} \right\}_+  \rho \right] \nonumber
\ee
where $\{ \, , \, \}_+$ denotes the anti-commutator. 
Notably for $m=1$ this relation produces the correct average work inline with Eq.~(\ref{1law}).  For $m=2$ it produces the same second moment as that of the power-based distribution $p(w)$, of the form (\ref{moments}). However, unless the Hamiltonians commute, i.e. $[H_{H}(\tau),H_{H}(0)]=0$, higher work moments arising from the Margenau Hill distribution will differ from higher moments for the continuous histories distribution, given in Eq.~(\ref{moments}). One consequence of this difference is that the average exponentiated work for the Margenau-Hill distribution does produce the standard Jarzynski equality \cite{Allahverdyan2014c}.

The quasi-probability distribution $p^{\MH}(w)$ also shares the time-reversal symmetry of the classical work distribution for a closed system, reflecting the fact that the dynamics must be reversible. This  follows straightforwardly with the same argument as in (\ref{eq:reverse}). We thus conclude that the alternative two-point quasi-probability $p^{\MH}(w)$ is also consistent with our desired conditions (i) and (ii), whilst leading to an alternative set of higher order moments than the set Eq.~(\ref{moments}). In the classical limit in which $\rho$, $H(0)$ and $H_{H}(\tau)$ commute, the moments associated with both $p^{\MH}(w)$ and $p(w)$ in the continuum limit $\Delta t \to 0$ converge to the moments given by the two-projective measurement distribution, $p^\tpm (w)$. In turn this distribution becomes the classical work probability distribution \cite{Jarzynski2015a}. To summarise our findings the properties of the four different distributions for work discussed in this paper are presented in Table~\ref{tab:compare}. 

\section{Discussion}\label{sec:discussion}

While in classical thermodynamics the act of measurement leaves no influence on the statistical properties of the system, in quantum mechanics one faces a dilemma: either one measures the fluctuating work and accepts disturbing the system, or one does not disturb the system but then there is no data to read off the fluctuating work value. Here we showed that it is however possible to define the alternative work distribution $p(w)$ associated with the latter {\it unmeasured} dynamics. Constructing the unmeasured work distribution is of interest as it allows the comparison with the measured, TPM work distribution. While in the classical limit both distributions are the same, in the quantum case it is insightful to  discuss their physical differences. 

To determine $p(w)$ we introduced the histories approach to quantum thermodynamics, which allows the construction of operator trajectories analogous to classical trajectories in phase space. We found that in contrast to the TPM distribution, the work distribution $p(w)$ fulfils energy conservation and time-reversal symmetry for general initial states and unitary evolutions. Moreover, for $p(w)$ the Jarzynski equality is not fulfilled in the continuum limit $\Delta t \to 0$, contrary to the work distribution arising from the TPM scheme. This deviation arises solely due to non-commutativity of Hamiltonians at different times.

As we demonstrated for an example in Section \ref{sec:example}, the resulting $p(w)$ is not positive for all closed dynamics and is thus a quasi-probability distribution. The emergence of negative probabilities highlights a distinctly non-classical feature of quantum work, reminiscent of other quasi-probabilities found in quantum mechanics, such as the Wigner function and Dirac distributions \cite{Lee1995,Dirac1945}. The negativity of the quasi-probability $p_{\vec{n}}$, Eq.~(\ref{linear}), for general operator trajectories has been shown to lead to violations of the Leggett-Garg inequality  \cite{Halliwell2015a} which indicates the failure of (classical) macro-realism. The negativity of $p(w)$ implies that the fluctuating work values $w$ cannot correspond to  outcomes of measurements - if they did then $p(w)$ must be a positive probability distribution. 

Within the histories approach we first used, analogously to the classical work definition, power trajectories that are integrated over time to construct $p(w)$ in Eq.~(\ref{workdist}). We secondly constructed a work distribution $p^{\MH}(w)$ in Eq.~(\ref{mh}) that identifies work with the energy change between end and start points of energy trajectories. Naively one may expect these two distributions to be the same in the continuum limit $\Delta t \to 0$, i.e. that the work for closed dynamics can be found by either integrating along the trajectories or by just considering the end points. Surprisingly however, in general the two distributions have different moments thus there is a difference between fluctuating energy change (end points) and fluctuating work (trajectory) even in an unmeasured closed quantum process. This shows that the way in which one defines quantum work has physical consequences for thermodynamic relations, such as the Jarzynski equality. 

Experimentally, quasi-probabilities commonly emerge in weak measurement schemes. In particular, the quasi-probability $p^{\MH}(w)$ can be obtained through weakly coupling the system to a pair of external detectors \cite{Lundeen2012,Bamber2014}. We expect that similar weak measurement schemes, such as \cite{Bednorz2012b,Solinas2015}, can be adapted to measure the continuous closed dynamics described by $p(w)$, and detect violations of the Jarzynski equality, see Eq.~(\ref{jarz}). The presented histories approach to quantum thermodynamics provides new insight into the non-classical properties of work in the quantum regime. The continuous trajectories picture offers comparison with the two-time TPM approach and highlights the effect of quantum measurement on the laws of thermodynamics.  

Future research will be required to construct an extension of the developed work concept to open systems and to analyse detailed fluctuation relations, such as Crooks' relation \cite{Crooks}. Moreover, it would be insightful to understand the link between the work definition based on the continuous operator trajectories used in the histories approach and other continuous quantum trajectory approaches. For example, for open quantum systems the microscopic {\it quantum trajectories} picture ``unravels''  the master equation into an ensemble of stochastically occurring trajectories \cite{Carmichael2009,Wiseman2010a}. Only recently has the quantum trajectories approach been used to identify  fluctuating work as well as quantum heat contributions \cite{Horowitz2013,Elouard2015,Alonso2015a}, while non-equilibrium entropy production has been discussed in \cite{Horowitz2013}. The classical correspondence could further be elucidated \cite{Jarzynski2015a} by translating the operator trajectories into the path integral formulation which relies on summing over all trajectories in phase space \cite{Feynman1963}.

\ack
The authors thank Paolo Solinas, Andrew Smith, Saar Rahav, Christopher Jarzynski and Simon Horsley for stimulating and insightful discussions. HM is supported by EPSRC through a Doctoral Training Grant. JA acknowledges support from EPSRC, grant EP/M009165/1, and the Royal Society. This research was supported by the COST network MP1209 ``Thermodynamics in the quantum regime''.

\

\bibliographystyle{unsrt}
\bibliography{historiesbib}

\begin{thebibliography}{10}

\bibitem{Mermin1985}
N.~Mermin.
\newblock {Is the moon there when nobody looks? Reality and the quantum
  theory}.
\newblock {\em Physics Today}, 38:38, 1985.

\bibitem{Jarzynski1997d}
C~Jarzynski.
\newblock {Nonequilibrium equality for free energy differences.}
\newblock {\em Phys. Rev. Lett.}, 78:2690, 1997.

\bibitem{Crooks}
Gavin Crooks.
\newblock {Entropy production fluctuation theorem and the nonequilibrium work
  relation for free energy differences.}
\newblock {\em Phys. Rev. E}, 60:2721, 1999.

\bibitem{Kawai2007a}
R~Kawai, J~M~R Parrondo, and C~{Van den Broeck}.
\newblock {Dissipation: the phase-space perspective.}
\newblock {\em Phys. Rev. Lett.}, 98:080602, 2007.

\bibitem{Seifert2012}
Udo Seifert.
\newblock {Stochastic thermodynamics, fluctuation theorems and molecular
  machines.}
\newblock {\em Rep. Prog. Phys}, 75:126001, 2012.

\bibitem{Vinjanampathy2015a}
Sai Vinjanampathy and Janet Anders.
\newblock {Quantum Thermodynamics.}
\newblock {\em Cont. Phys.}, 7514:1, 2015.

\bibitem{Allahverdyan2005a}
A~E Allahverdyan and Th~M Nieuwenhuizen.
\newblock {Fluctuations of work from quantum subensembles: the case against
  quantum work-fluctuation theorems.}
\newblock {\em Phys. Rev. E}, 71:066102, 2005.

\bibitem{Talkner2007c}
Peter Talkner, Eric Lutz, and Peter H{\"{a}}nggi.
\newblock {Fluctuation theorems: Work is not an observable.}
\newblock {\em Phys. Rev. E}, 75:050102, 2007.

\bibitem{Esposito2009}
Massimiliano Esposito, Upendra Harbola, and Shaul Mukamel.
\newblock {Nonequilibrium fluctuations, fluctuation theorems, and counting
  statistics in quantum systems.}
\newblock {\em Rev. Mod. Phys.}, 81:1665, 2009.

\bibitem{Allahverdyan2014c}
A~E Allahverdyan.
\newblock {Nonequilibrium quantum fluctuations of work.}
\newblock {\em Phys. Rev. E}, 90:032137, 2014.

\bibitem{Jarzynski2015a}
Christopher Jarzynski, H.~T. Quan, and Saar Rahav.
\newblock {Quantum-Classical Correspondence Principle for Work Distributions.}
\newblock {\em Phys. Rev. X}, 5:031038, 2015.

\bibitem{Solinas2015}
P~Solinas and S~Gasparinetti.
\newblock {Full distribution of work done on a quantum system for arbitrary
  initial states.}
\newblock {\em Phys. Rev. E}, 92:042150, 2015.

\bibitem{Alonso2015a}
Jose~Joaquin Alonso, Eric Lutz, and Alessandro Romito.
\newblock {Thermodynamics of Weakly Measured Quantum Systems}.
\newblock {\em Phys. Rev. Lett.}, 116:080403, 2016.

\bibitem{Deffner2016a}
Sebastian Deffner, Juan~Pablo Paz, and Wojciech~H. Zurek.
\newblock {Quantum work and the thermodynamic cost of quantum measurements.}
\newblock {\em Phys. Rev. E}, 94:1, 2016.

\bibitem{Perarnau-Llobet2016}
Mart{\'{i}} Perarnau-Llobet, Elisa B{\"{a}}umer, Karen {V. Hovhannisyan},
  Marcus Huber, and Antonio Acin.
\newblock {Quantum Fluctuations of Work and Generalised Quantum Measurements.}
\newblock {\em arXiv:1606.08368.}, 2016.

\bibitem{Hanggi2015}
Peter H{\"{a}}nggi and Peter Talkner.
\newblock {The other QFT.}
\newblock {\em Nat. Phys.}, 11:108, 2015.

\bibitem{Talkner2008b}
Peter Talkner, Peter H{\"{a}}nggi, and Manuel Morillo.
\newblock {Microcanonical quantum fluctuation theorems.}
\newblock {\em Phys. Rev. E}, 77:051131, 2008.

\bibitem{Talkner2007a}
Peter Talkner and Peter H{\"{a}}nggi.
\newblock {The Tasaki–Crooks quantum fluctuation theorem.}
\newblock {\em J. Phys. A}, 40:F569, 2007.

\bibitem{Kafri2012b}
Dvir Kafri and Sebastian Deffner.
\newblock {Holevo's bound from a general quantum fluctuation theorem.}
\newblock {\em Phys. Rev. A}, 86:044302, 2012.

\bibitem{Batalhao2014}
Tiago~B. Batalh{\~{a}}o, Alexandre~M. Souza, Laura Mazzola, Ruben Auccaise,
  Roberto~S. Sarthour, Ivan~S. Oliveira, John Goold, Gabriele {De Chiara},
  Mauro Paternostro, and Roberto~M. Serra.
\newblock {Experimental Reconstruction of Work Distribution and Study of
  Fluctuation Relations in a Closed Quantum System.}
\newblock {\em Phys. Rev. Lett.}, 113:140601, 2014.

\bibitem{an2015}
Shuoming An, Jing-Ning Zhang, Mark Um, Dingshun Lv, Yao Lu, Junhua Zhang,
  Zhang-Qi Yin, H.~T. Quan, and Kihwan Kim.
\newblock {Experimental test of the quantum Jarzynski equality with a
  trapped-ion system.}
\newblock {\em Nat. Phy.}, 11:193, 2014.

\bibitem{Deffner2011a}
Sebastian Deffner and Eric Lutz.
\newblock {Nonequilibrium entropy production for open quantum systems.}
\newblock {\em Phys. Rev. Lett.}, 107:140404, 2011.

\bibitem{Goold2015b}
John Goold, Mauro Paternostro, and Kavan Modi.
\newblock {Nonequilibrium Quantum Landauer Principle.}
\newblock {\em Phys. Rev. Lett.}, 114:060602, 2015.

\bibitem{Talkner2009}
Peter Talkner, Michele Campisi, and Peter H{\"{a}}nggi.
\newblock {Fluctuation theorems in driven open quantum systems.}
\newblock {\em J. Stat. Mech.}, 2009:P02025, 2009.

\bibitem{Kammerlander2016}
P.~Kammerlander and J.~Anders.
\newblock {Coherence and measurement in quantum thermodynamics.}
\newblock {\em Sci. Rep.}, 6:22174, 2016.

\bibitem{Griffiths1984}
Robert~B. Griffiths.
\newblock {Consistent histories and the interpretation of quantum mechanics.}
\newblock {\em J. Stat. Phys.}, 36:219, 1984.

\bibitem{Gell-Mann1993a}
Murray Gell-Mann and James~B. Hartle.
\newblock {Classical equations for quantum systems.}
\newblock {\em Phys. Rev. D}, 47:3345, 1993.

\bibitem{HALLIWELL1995}
J.~J. Halliwell.
\newblock {A Review of the Decoherent Histories Approach to Quantum Mechanics.}
\newblock {\em Annals N.Y. Acad. Sci.}, 755:726, 1995.

\bibitem{Goldstein1995}
Sheldon Goldstein and Don~N. Page.
\newblock {Linearly positive histories: Probabilities for a robust family of
  sequences of quantum events.}
\newblock {\em Phys. Rev. Lett.}, 74:3715, 1995.

\bibitem{Hartle2004}
James~B. Hartle.
\newblock {Linear positivity and virtual probability.}
\newblock {\em Phys. Rev. A}, 70:022104, 2004.

\bibitem{Solinas2013}
Paolo Solinas, Dmitri~V. Averin, and Jukka~P. Pekola.
\newblock {Work and its fluctuations in a driven quantum system.}
\newblock {\em Phys. Rev. B}, 87:060508, 2013.

\bibitem{PrasannaVenkatesh2015a}
B~{Prasanna Venkatesh}, Gentaro Watanabe, and Peter Talkner.
\newblock {Quantum fluctuation theorems and power measurements.}
\newblock {\em N. J. Phys}, 17:075018, 2015.

\bibitem{Subas2012}
Y~Subasi and B~L Hu.
\newblock {Quantum and classical fluctuation theorems from a decoherent
  histories, open-system analysis.}
\newblock {\em Phys. Rev. E}, 85:011112, 2012.

\bibitem{Li2013}
Huanan Li and J.S. Wang.
\newblock {Work distribution under continuous quantum histories.}
\newblock {\em arXiv:1304.6286.}, 2013.

\bibitem{Elouard2015}
C.~Elouard, Alexia Auff{\`{e}}ves, and M.~Clusel.
\newblock {Stochastic thermodynamics in the quantum regime.}
\newblock {\em arXiv:1507.00312.}, 2015.

\bibitem{Johansen2007}
Lars~M. Johansen.
\newblock {Quantum theory of successive projective measurements.}
\newblock {\em Phys. Rev. A}, 76:012119, 2007.

\bibitem{Yi2013b}
Juyeon Yi and Beom~Jun Kim.
\newblock {Thermodynamic arrow of time of quantum projective measurements.}
\newblock {\em EPL}, 103:20006, 2013.

\bibitem{Lostaglio2015b}
Matteo Lostaglio, Kamil Korzekwa, David Jennings, and Terry Rudolph.
\newblock {Quantum Coherence, Time-Translation Symmetry, and Thermodynamics.}
\newblock {\em Phys. Rev. X}, 5:021001, 2015.

\bibitem{Hartle2008}
James~B. Hartle.
\newblock {Quantum mechanics with extended probabilities.}
\newblock {\em Phys. Rev. A}, 78:012108, 2008.

\bibitem{Gell-Mann2012}
Murray Gell-Mann and James~B. Hartle.
\newblock {Decoherent histories quantum mechanics with one real fine-grained
  history}.
\newblock {\em Phys. Rev. A}, 85:062120, 2012.

\bibitem{Lee1995}
Hai-Woong Lee.
\newblock {Theory and application of the quantum phase-space distribution
  functions.}
\newblock {\em Phys. Rep.}, 259:147, 1995.

\bibitem{Garraway1993}
B~M Garraway and P~L Knight.
\newblock {Quantum superpositions, phase distributions and
  quasi-probabilities.}
\newblock {\em Phys. Script.}, T48:66, 1993.

\bibitem{Banaszek1999}
K~Banaszek, C~Radzewicz, K~W{\'{o}}dkiewicz, and J~S Krasinski.
\newblock {Direct measurement of the Wigner function by photon counting.}
\newblock {\em Phys. Rev. A}, 60:674, 1999.

\bibitem{Lundeen2012}
Jeff~S Lundeen and Charles Bamber.
\newblock {Procedure for direct measurement of general quantum states using
  weak measurement.}
\newblock {\em Phys. Rev. Lett.}, 108:070402, 2012.

\bibitem{Bamber2014}
Charles Bamber and Jeff~S Lundeen.
\newblock {Observing Dirac's classical phase space analog to the quantum
  state.}
\newblock {\em Phys. Rev. Lett.}, 112:070405, 2014.

\bibitem{Cohen1982}
Joel~E. Cohen, Shmuel Friedland, Tosio Kato, and Frank~P. Kelly.
\newblock {Eigenvalue inequalities for products of matrix exponentials}.
\newblock {\em Lin. Alg. App.}, 45, 1982.

\bibitem{Aberg2016}
Johan Aberg.
\newblock {Fully quantum fluctuation theorems.}
\newblock {\em arXiv:1601.01302.}, 2016.

\bibitem{Chernyak2004}
Vladimir Chernyak and Shaul Mukamel.
\newblock {Effect of quantum collapse on the distribution of work in driven
  single molecules.}
\newblock {\em Phys. Rev. Lett.}, 93:048302, 2004.

\bibitem{Margenau1961}
Henry Margenau and Robert~Nyden Hill.
\newblock {Correlation between Measurements in Quantum Theory.}
\newblock {\em Prog. Theo. Phys.}, 26:722, 1961.

\bibitem{Dirac1945}
P.~A.~M. Dirac.
\newblock {On the Analogy Between Classical and Quantum Mechanics.}
\newblock {\em Rev. Mod. Phys.}, 17:195, 1945.

\bibitem{Halliwell2015a}
J.~J. Halliwell.
\newblock {Leggett-Garg inequalities and no-signaling in time: A
  quasiprobability approach.}
\newblock {\em Phys. Rev. A}, 93:022123, 2016.

\bibitem{Bednorz2012b}
Adam Bednorz, Wolfgang Belzig, and Abraham Nitzan.
\newblock {Nonclassical time correlation functions in continuous quantum
  measurement}.
\newblock {\em N. J. Phys}, 14:013009, 2012.

\bibitem{Carmichael2009}
Howard~J. Carmichael.
\newblock {\em {Statistical Methods in Quantum Optics 2: Non-Classical
  Fields}}.
\newblock Springer, Berlin, Heidelberg, 2009.

\bibitem{Wiseman2010a}
Howard Wiseman and Gerard~J. Milburn.
\newblock {\em {Quantum Measurement and Control}}.
\newblock Cambridge University Press, 2010.

\bibitem{Horowitz2013}
Jordan~M Horowitz and Juan M~R Parrondo.
\newblock {Entropy production along nonequilibrium quantum jump trajectories.}
\newblock {\em N. J. Phys}, 15:085028, 2013.

\bibitem{Feynman1963}
R.~P. Feynman and F.~L. Vernon.
\newblock {The theory of a general quantum system interacting with a linear
  dissipative system.}
\newblock {\em Ann. Phys.}, 24:118, 1963.

\end{thebibliography}

\end{document}